\documentclass[12pt]{article}
\usepackage{amssymb}

\input{tcilatex}
\begin{document}

\title{Size limiting in Tsallis statistics.}
\author{Hari M. Gupta, Jos\'{e} R. Campanha, Sidney J. Schinaider \\
IGCE - Departamento de F\'{\i}sica, IGCE, UNESP\\
Caixa Postal 178, CEP 13500-970\\
Rio Claro - S\~{a}o Paulo - Brazil}
\maketitle

\begin{abstract}
Power law scaling is observed in many physical, biological and
socio-economical complex systems and is now considered as an important
property of these systems. In general, power law exists in the central part
of the distribution. It has deviations from power law for very small and
very large step sizes. Tsallis, through non-extensive thermodynamics,
explained power law distribution in many cases including deviation from the
power law, both for small and very large steps. In case of very large steps,
they used heuristic crossover approach.

In real systems, the size is limited and thus, the size limiting factor is
important. In the present work, we present an alternative model in which we
consider that the entropy factor q decreases with step size due to the
softening of long range interactions or memory. This explains the deviation
of power law for very large step sizes. Finally, we apply this model for
distribution of citation index of scientists and examination scores and are
able to explain the entire distribution including deviations from power law.
\end{abstract}

\textbf{I. Introduction}

\bigskip

\bigskip

Only recently physicists started to study the natural systems as a whole
rather than in parts [1-6] and are interested in holistic properties of
these systems normally called \textquotedblleft Complex
Systems\textquotedblright . These systems are difficult to understand from
the basic principles. The difficulties in understanding these systems arise
from the fact that, in most of the cases, a large number of elementary
interactions is taking place at the same time for a large number of
components. Further, these systems are in constant evolution and do not have
an equilibrium state [1]. Power law scaling [7,8] is observed in many
biological [9-11], physical [2,12-20] and socio-economical complex systems
[21-29] and it is now considered an important property of them. As
socio-economical systems also have almost the same characteristics,
physicists are also studying some of these systems.

In 1988, Tsallis [30] presented non-extensive thermodynamics in which he
incorporated long range interactions and long memory effects. He proposed a
generalized definition of entropy $(S_{q})$:

\bigskip 
\begin{equation}
S_{q}=C\frac{1-\sum_{i=1}^{W}p_{i}^{q}}{q-1}
\end{equation}

\[
(\sum_{i=1}^{W}p_{i}=1) 
\]

\bigskip

\noindent where $C$ is a positive constant, and $W$ is the total number of
microscopic possibilities of the system. $q$ is an entropic index, which
plays a central role and is related to long range interactions and long
memory effect in a network. This expression recovers the usual
Boltzmann-Gibbs entropy $(-C\sum_{i=1}^{W}p_{i}\ln p_{i})$ in the limit $%
q\rightarrow 1$, i.e. in short range interactions [31]. The size frequency
distribution function $N(x)$ is given through

\bigskip

\begin{equation}
\frac{dN(x)}{dx}=-\lambda N(x)
\end{equation}

\bigskip

\noindent where $\lambda $ is a positive constant. $N(x)$ is the frequency
probability of step size $x$.

This gives

\bigskip

\begin{equation}
N(x)=N_{0}\exp (-\lambda x)
\end{equation}

\bigskip

\noindent where $N_{0}$ is a normalization constant, thus we have an
exponential decay which is exactly the case of Boltzmann statistics
considering short range interactions. However for $q>1$, a more generalized
equation [32] holds giving:

\bigskip

\begin{equation}
\frac{dN(x)}{dx}=-\lambda N^{q}(x)
\end{equation}

\bigskip

\noindent hence

\bigskip

\begin{equation}
N(x)=\frac{N_{0}}{[1+(q-1)\lambda x]^{\frac{1}{q-1}}}
\end{equation}

\bigskip

\noindent or in an alternative form:

\bigskip

\begin{equation}
N(x)=\frac{N_{0}}{[1+\beta x]^{\alpha }}
\end{equation}

\bigskip

\noindent where $\beta =(q-1)\lambda $ and $\alpha =1/(q-1).$

For relatively large values of x, the distribution becomes

\bigskip

\begin{equation}
N(x)\approx const.x^{-\alpha }
\end{equation}

\bigskip

\noindent i.e. a power law. In this case, a $logN(x)vs.log(x)$ plot exhibits
a straight line for large values of $x$. Power law distribution can not
continue forever in real systems. It has to be truncated in some way to
avoid infinite variance.

Recently, we have shown that by gradually truncating power law distribution
after certain critical value, we are able to explain the entire distribution
including very large steps in financial and physical complex systems
[33-35]. Although this model explains empirical results for large step
sizes, it has an undesirable discontinuity at critical step size. Also, this
model fails for small steps.

In discussing folding-unfolding phenomena that occurs in proteins, Tsallis
et. al. [36] argued that with the increase of temperature, increases thermal
motion, which in turn decreases long memory or long range interactions and
finally decreases entropy index q. Thus q approaches to 1. At low
temperatures, the distribution function, which shows power law becomes
exponential at relatively high temperatures. For a fixed temperature, q is
considered to be a constant. In order to consider long range departure, they
assume a crossover to another type of behavior and modify Equation (4) as

\bigskip

\begin{equation}
\frac{dN(x)}{dx}=-\mu _{r}N^{r}(x)-(\lambda -\mu _{r})N^{q}(x)
\end{equation}

\bigskip

\noindent $\mu _{r}$ is very small compared to $\lambda $. That gives a
crossover between two different power laws (respectively characterized by $q$
and $r$) or from power law to normal distribution within a nonextensive
scenario, which is definitely a case for many complex systems and gives
multifractality. The cut-off is sharpest when $r=1$. In this case

\begin{equation}
N(x)=\frac{N_{0}}{(1-\frac{\lambda }{\mu _{1}}+\frac{\lambda }{\mu _{1}}%
e^{(q-1)\mu _{1}x})^{\frac{1}{q-1}}}
\end{equation}

\bigskip

Although cross-over behavior as suggested by Tsallis can avoid an infinite
variance, in the present work, we are looking for another possibility, i.e.
truncation of power law due to finite size in real systems which in fact is
not a cross-over behavior. We therefore suggest an alternative approach to
address the long time $(t)$ or long distance $(x)$ departures. We\ consider
that entropy factor $q$ decreases with step size $(x)$ due to the softening
of long range interactions or memory effects due to finite size in real
systems which arises because of physical limitation of the component or the
system itself. Thus q depends on the step size. This is similar as
anharmonic terms are important for calculating potential energy in lattice
vibrations.

Finally, we apply this model for the distribution of the citation index of
scientists and examination scores of an entrance examination and compare it
with Tsallis approach [36].

\bigskip

\textbf{II. The model}

\bigskip

\bigskip

The physical limiting factor is of a very small importance for small steps,
while it is necessary for larger steps. Entropy index $q$ is equal to $1$ in
the absence of long memory or long range interactions. Thus the information
about these interactions is given through $(q-1)$. We consider that this
factor approaches to zero for very large values of $x$ due to finite size in
real systems. In general, for this, we propose

\bigskip

\begin{equation}
(q(x)-1)=\frac{(q_{0}-1)}{1+\sum\limits_{j}\theta _{j}x^{j}}
\end{equation}

\bigskip

\noindent where $q_{0}$ and $q(x)$ are values of entropy index $q$ for step
size zero and step size $x$ respectively. $\theta _{i}$ and $i$ are
adjustable parameters, depending on the size limiting.

To simplify, we propose an exponential decay i.e$.$

\bigskip

\begin{equation}
q(x)-1=(q_{0}-1)\exp (-(\theta x)^{i})
\end{equation}

\bigskip

\noindent where $\theta $ and $i$ show the rate of decrease of the
importance of these interactions with the increase of step size $x.$ The
higher value of $i$ indicates a sharper cut-off.

For very large values of $x$, $q(x)$ approaches to $1$ and thus gives normal
distribution as required through central limit theorem. In the present model
the distribution function is given through:

\bigskip

\begin{equation}
N(x)=N_{0}[1+(q_{0}-1)\lambda x\exp (-(\theta x)^{i})]^{-(\exp ((\theta
x)^{i}))/(q_{0}-1)}
\end{equation}

\bigskip

In Figure 1 we compare $N(x)$ $vs.$ $x$ in Tsallis approach through Equation
(9) and present approach through Equation (12) for very large steps. Under
the present model, the gradual truncation of the power law can be adjusted
from very sharp to very slow through the value of $i$ without interfering in
power law behavior in the central part of the distribution. This is not
possible in Tsallis approach. For larger values of $\mu _{r}$ (line II of
Figure 1), we can get a sharp cut-off, but then it deviates significatory
from power law in the central part.

\bigskip

\[
\FRAME{itbpF}{13.0479cm}{10.0364cm}{0cm}{}{}{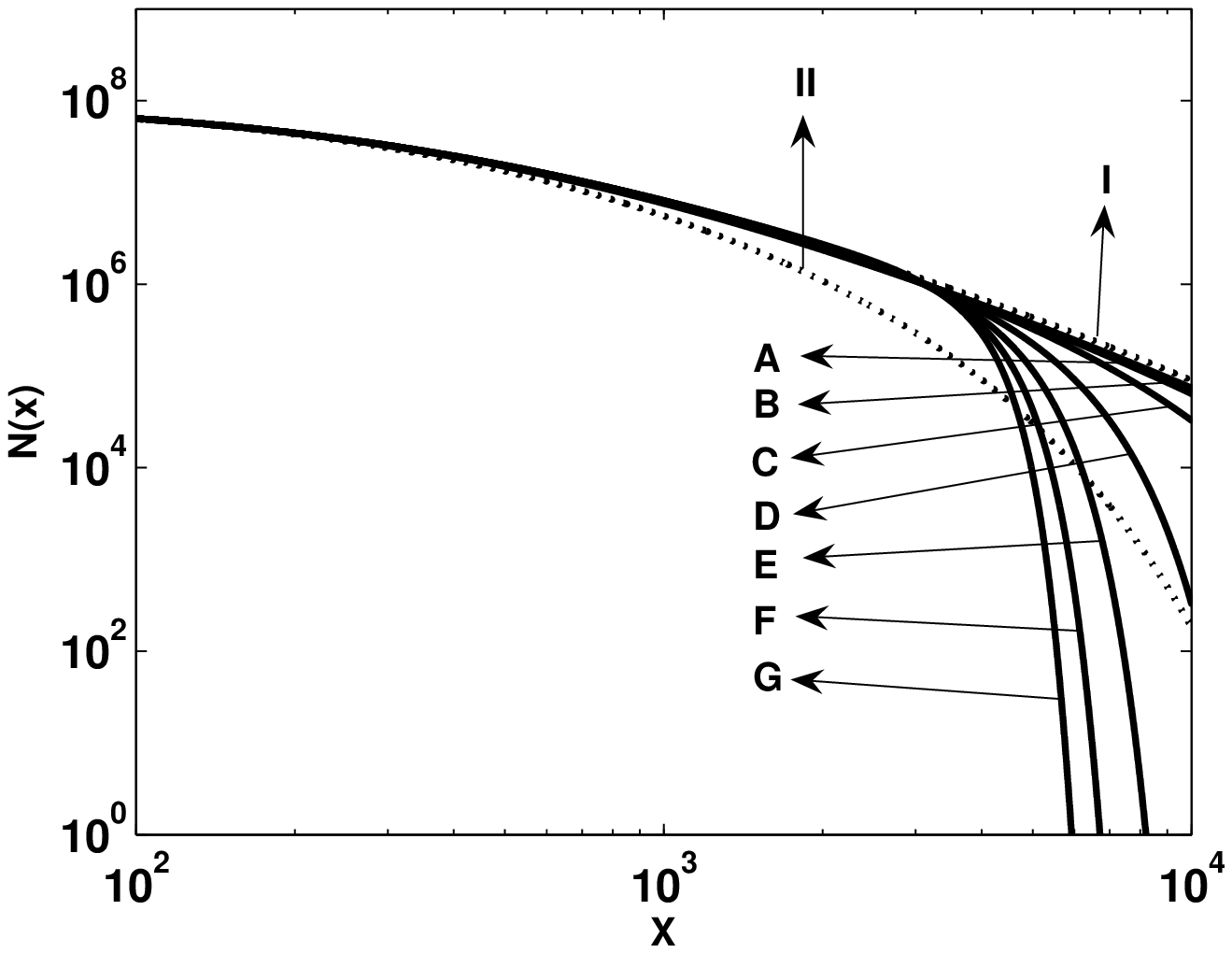}{\special%
{language "Scientific Word";type "GRAPHIC";maintain-aspect-ratio
TRUE;display "FULL";valid_file "F";width 13.0479cm;height 10.0364cm;depth
0cm;original-width 5.8219in;original-height 4.3708in;cropleft "0";croptop
"0.9494";cropright "0.9283";cropbottom "0";filename
'fig1_teorica.eps';file-properties "XNPEU";}}
\]

\textbf{Figure 1 }-- Theoretical distribution of Tsallis and present Model
in log-log scale. We consider: $N_{0}=1.10^{8}$; $\lambda =0.005$; $q_{0}=1.5
$. Curves I and II are through Tsallis model considering $r=1$ and $%
\mu
_{r}=1.10^{-4}$ and $1.10^{-3}$ respectively. Curves A, B, C, D, E, F, and G
are through present model considering $i=1/3$ and $\theta =3.10^{-7}$ (Curve
A), $i=1/2$ and $\theta =3.10^{-6}$ (Curve B), $i=1$ and $\theta =$ $%
3.10^{-5}$ (Curve C), $i=2$ and $\theta =$ $1.10^{-4}$ (Curve D), $i=3$ and $%
\theta =$ $1.5.10^{-4}$ (Curve E), $i=4$ and $\theta =$ $1.8.10^{-4}$ (Curve
F) and $i=5$ and $\theta =$ $2.10^{-4}$ for Curve G.

\bigskip

\bigskip \textbf{III. Distribution of citation Index}

\bigskip

Now we apply this model to describe the distribution of citation index of
the scientists. The citation index of a scientist is the total number of
times that his articles are cited in other articles. The citation patterns
of scientific publications form a rather complex network [37]. Here nodes
are published papers. The citation of an article is an interaction of a
scientific work with other scientific works. Most of the articles are cited
in the proper group only. However many articles go beyond it and are of the
interest of others. Some pioneer articles are cited for many decades. Thus,
citation index arises from both short and long range interactions and can be
treated through statistical distribution based on Tsallis entropy concept.

The fact that a scientist is cited more times facilitates him to get more
financial help to his research projects and better students. Some other
small groups also came in his influence. This, in turn, contributes to form
a better and larger group. In physical terms these effects produce long
range interactions. A pioneer work is also cited just to complete
introduction of a problem, and thus is cited for a larger time [42],
although the problem is not directly connected to it. This gives a long
range memory

In case of the scientist's citation index, unfortunately, reliable
information is available only for some of the most cited physicists or
chemists. There are many scientists with the same name and do not exist a
rigid control to separate them. This can make a significant error for low
cited scientists. Thus, it is not possible to have a complete statistical
analysis as we did in the case of scientific publications [38] and found
that non-extensive thermodynamical distribution (Tsallis statistics) is
valid over eight orders of magnitude $(10^{-4}$ to $10^{4}).$ It is
therefore interesting to construct a Zipf plot [39] in case of citation
index of scientists, in which the number of citations of the $n^{th}$ most
cited scientists out of an ensemble of $M$ scientists is plotted versus rank 
$n$. By its very definition, the Zipf plot is closely related to the
cumulative large $x$ tail of the citation distribution. This plot is
therefore well suited for determining the large $x$ of the citation
distribution. This plot also smoothes out the fluctuations in the
high-citation tail and thus facilitates quantitative analysis.

Given an ensemble of M scientists and the corresponding number of citations
for each of these scientists in rank order $Y_{1}\geqslant Y_{2}\geqslant
Y_{3}...\geqslant Y_{n}\geqslant ...Y_{M}$, then the number of citations of
the $n^{th}$ most-cited scientist $Y_{n}$ may be estimated by the criterion
[39]:

\bigskip

\begin{equation}
\int_{Y_{n}}^{\infty }N(x)dx=n
\end{equation}

\bigskip

This specifies that there are $n$ scientists out of the ensemble of $M$
which are cited at least $Y_{n}$ times. From the dependence of $Y_{n}$ on $n$
in a Zipf plot, one can test whether it agrees with a hypothesized form for $%
N(x)$.

We analyze citation index of (a) the most-cited Brazilian physicists and
chemists and (b) Internationally most cited physicists and chemists. By
Brazilian scientists we mean all scientists who are working in Brazil or
have a permanent working address in Brazil. All physicists (chemists)
including Brazilian physicists publish their work in the same Journals and
work almost on the same problems. Physics, like any other basic science, is
the same all over the world. In case of the internationally most cited
physicists, we have distribution function only for a few physicists ($%
\approx 1000$). Considering the most cited Brazilian physicists with the
same parameters, we widely extend this range to roughly 100.000 most cited
physicists. Thus the distribution of citation index of scientists is more
critically discussed.

In Figure 2 we plot citation number $(Y_{n})$ versus rank $(n)$ for first
205 Brazilian physicists in 1999 [40] and compare with Tsallis and the
present model. For the present model we use the following parameters $%
N_{0}=1.5,$ $q_{0}=1.39,$ $\lambda =0.0055$ and $\theta =2.5.10^{-4}$ and $%
i=1$. For Tsallis statistics (Equation 9), the parameters used are $%
N_{0}=1.65,$ $q_{0}=1.39,$ $\lambda =0.0055,$ $\mu _{r}=3.10^{-4}$ and $r=1$.

\bigskip

\[
\FRAME{itbpF}{12.9645cm}{9.127cm}{0cm}{}{}{Figure}{\special{language
"Scientific Word";type "GRAPHIC";maintain-aspect-ratio TRUE;display
"FULL";valid_file "T";width 12.9645cm;height 9.127cm;depth
0cm;original-width 11.2798in;original-height 7.9208in;cropleft "0";croptop
"1";cropright "1";cropbottom "0";tempfilename
'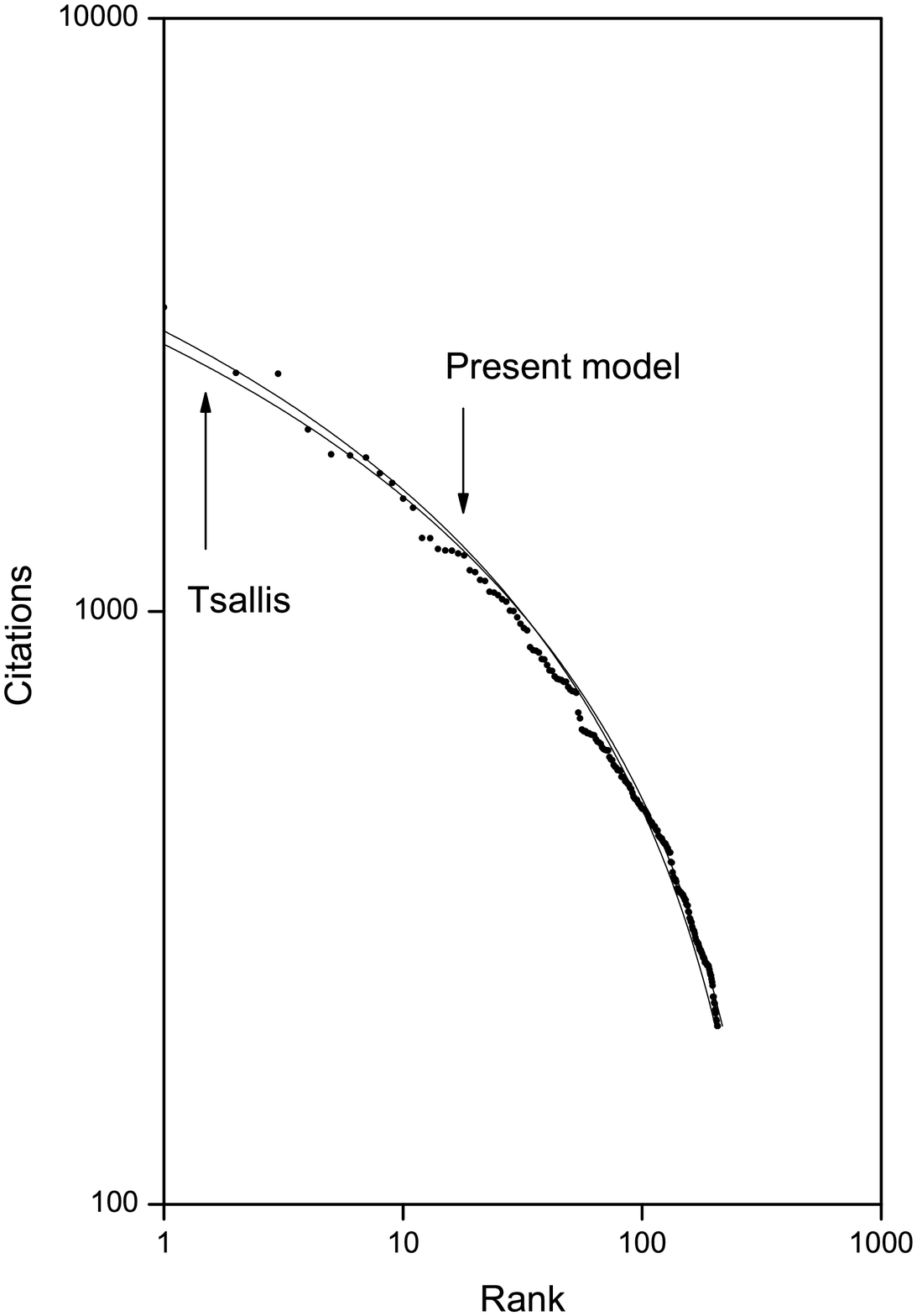';tempfile-properties "XPR";}}
\]

\begin{center}
\textbf{Figure 2} Zipf plot of the number of citation of the $n^{th}$ ranked
Brazilian

physicist $Y_{n}$ versus rank n on a double logarithmic scale.

\bigskip
\end{center}

\[
\FRAME{itbpF}{13.0479cm}{10.0364cm}{0cm}{}{}{Figure}{\special{language
"Scientific Word";type "GRAPHIC";maintain-aspect-ratio TRUE;display
"FULL";valid_file "T";width 13.0479cm;height 10.0364cm;depth
0cm;original-width 11.2798in;original-height 7.9208in;cropleft "0";croptop
"1.0149";cropright "0.9283";cropbottom "0";tempfilename
'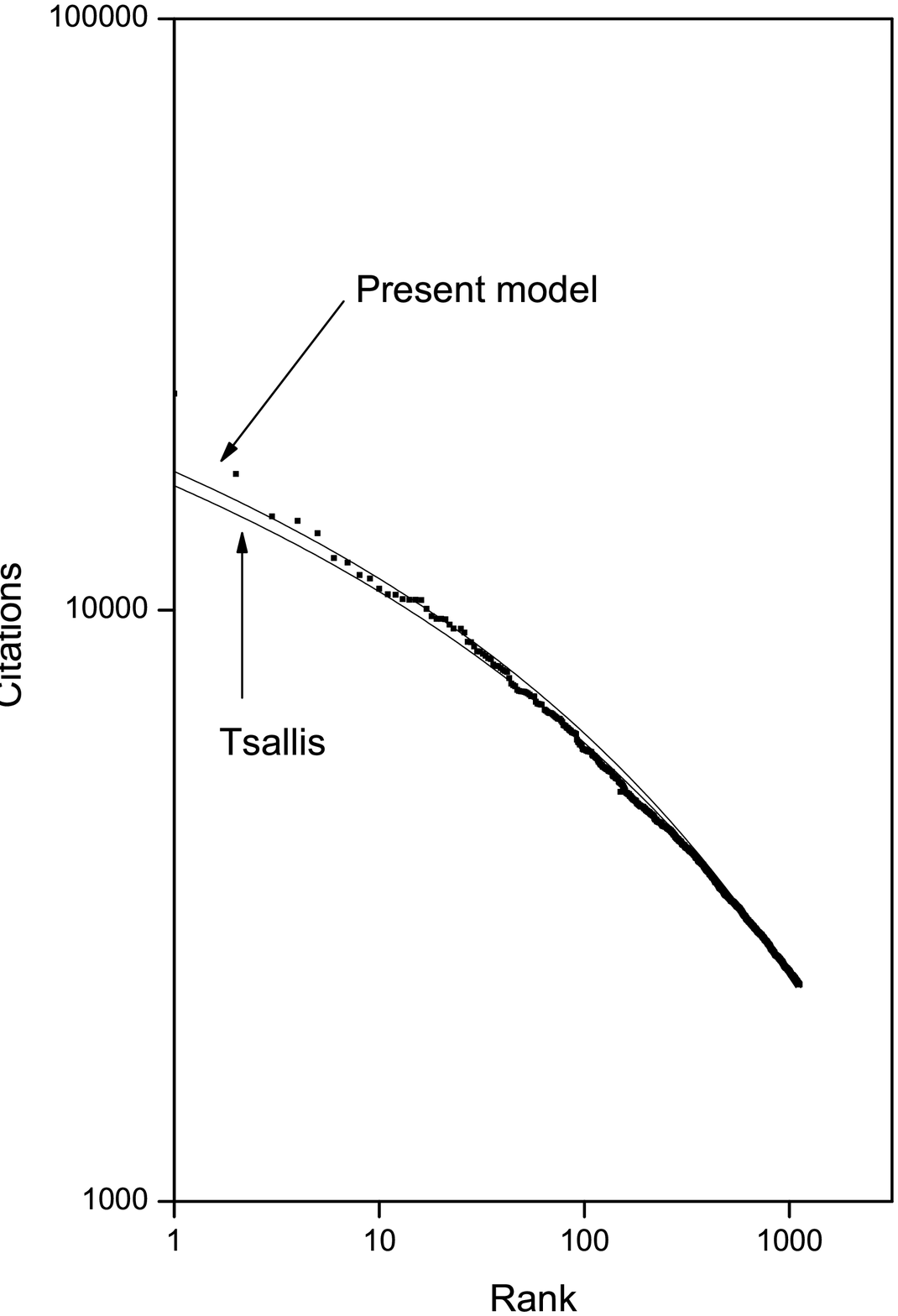';tempfile-properties "XPR";}}
\]%
\bigskip 

\begin{center}
\textbf{Figure 3} Zipf plot of the number of citation of the $n^{th}$ ranked
International physicist $Y_{n}$ versus rank n on a double logarithmic scale.
\end{center}

\bigskip

In Figure 3, we plot citation number $(Y_{n})$ versus rank $(n)$ for 1120
most cited physicists over the period 1981-June 1997 [41] and compare it
with the theoretical curve with the same value of q$_{0}$ and $\lambda $\ as
used in Figure 2. We changed the value of constant $N_{0}$ from $1.5$ to $95$
as total number of physicists is much larger in this case. We are
considering Brazilian physicists citations roughly $1.5\%$ of the total
citations, which is reasonable. The value of $\theta $ changes from $%
2.5.10^{-4}$ to $3.10^{-5}$. This shows that size limiting factors are much
more important for Brazilian physicists compared to internationally most
cited physicists which are mostly from U.S.A. This is perhaps due to the
absence of basic infrastructure and large research laboratories in Brazil to
work on important problems particularly in experimental physics. It is
interesting to note that 8 out of 10 most cited Brazilian physicists are
working in theoretical physics.\ In case of Tsallis statistics we use the
same basic parameters ( $q$ and $\lambda $\ ) as in Figure 2. We change $%
N_{0}$ from $1.65$ to $135$ and $\mu $ from $3.10^{-4}$ to $3.8.10^{-4}$. We
again observe a good agreement both in ours and Tsallis model. Note that we
are able to explain both distributions with the same values of basic
parameters.

\bigskip \bigskip 
\[
\FRAME{itbpF}{13.0479cm}{10.0364cm}{0cm}{}{}{Figure}{\special{language
"Scientific Word";type "GRAPHIC";maintain-aspect-ratio TRUE;display
"FULL";valid_file "T";width 13.0479cm;height 10.0364cm;depth
0cm;original-width 11.2798in;original-height 7.9208in;cropleft "0";croptop
"1.0149";cropright "0.9283";cropbottom "0";tempfilename
'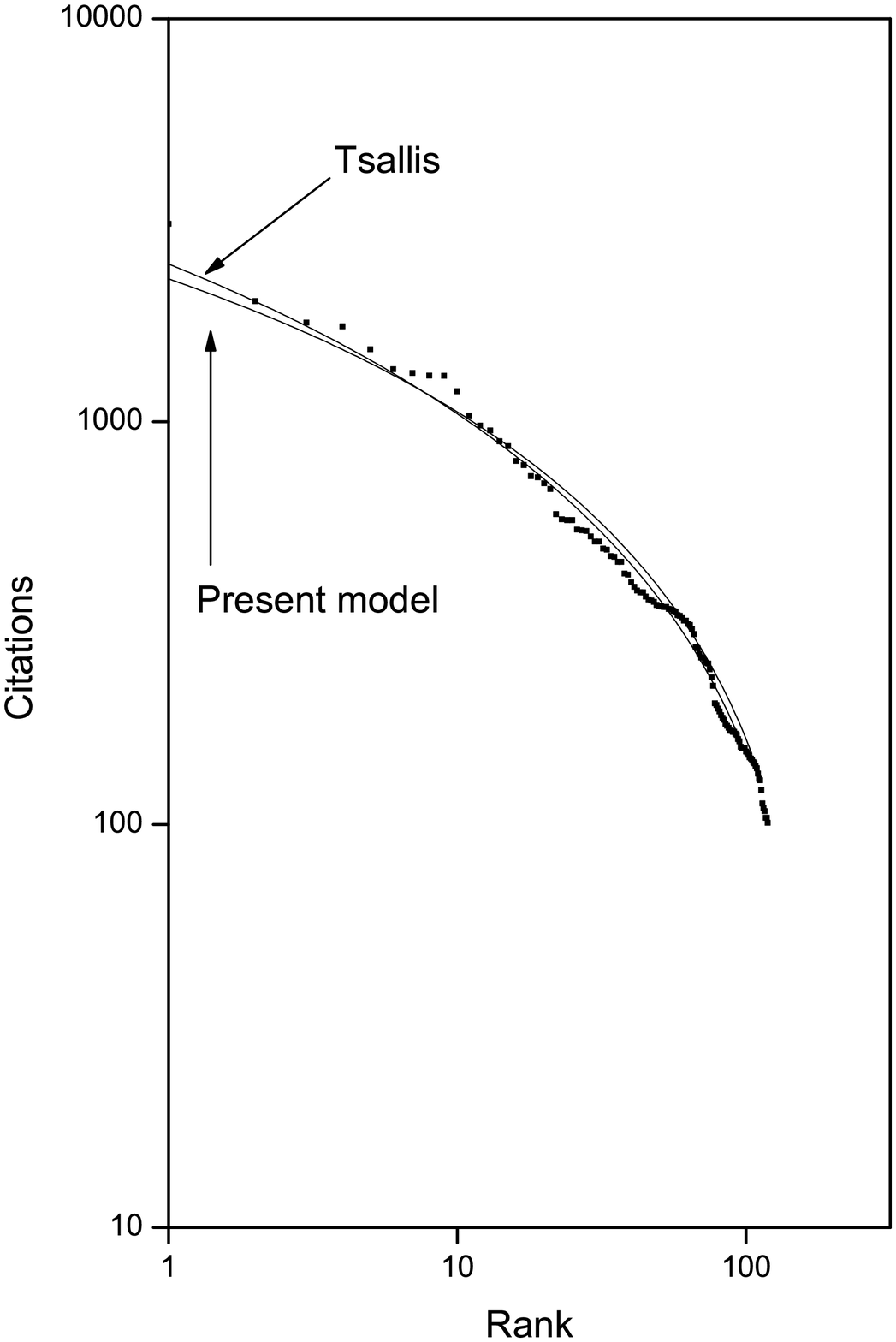';tempfile-properties "XPR";}}
\]

\begin{center}
\textbf{Figure 4 }Zipf plot of the number of citation of the $n^{th}$ ranked
Brazilian

chemist $Y_{n}$ versus rank n on a double logarithmic scale.

\bigskip
\end{center}

\bigskip

\[
\FRAME{itbpF}{13.0479cm}{10.0364cm}{-0.0527cm}{}{}{Figure}{\special{language
"Scientific Word";type "GRAPHIC";maintain-aspect-ratio TRUE;display
"FULL";valid_file "T";width 13.0479cm;height 10.0364cm;depth
-0.0527cm;original-width 11.3584in;original-height 7.9329in;cropleft
"0";croptop "1.0204";cropright "0.9283";cropbottom "0";tempfilename
'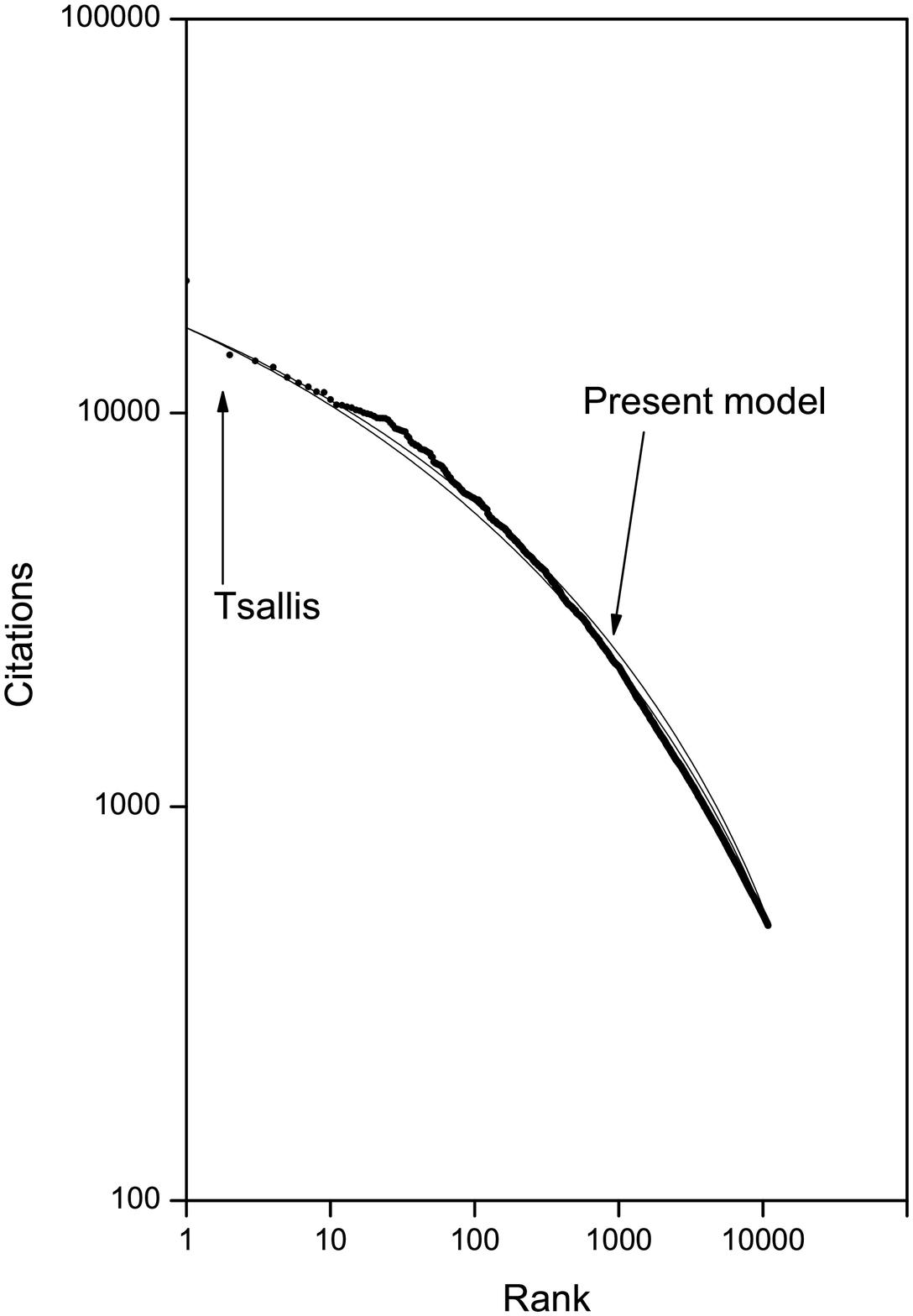';tempfile-properties "XPR";}}
\]

\begin{center}
\textbf{Figure 5} Zipf plot of the number of citation of the $n^{th}$ ranked
International chemist $Y_{n}$ versus rank n on a double logarithmic scale.
\end{center}

\bigskip

In Figure 4, we plot citation number $(Y_{n})$ versus rank $(n)$ for the
first 119 Brazilian chemists in 1999 [40]. We compare it with present model
and Tsallis model. We consider following parameters for the present model $%
N_{0}=0.8,$ $q_{0}=1.35,$ $\lambda =0.006$ and $\theta =2.2.10^{-4}$ and $i=1
$. For Tsallis statistics we use $N_{0}=0.8,$ $q_{0}=1.35,$ $\lambda =0.006$
and $\mu _{r}=1.10^{-3}$ and $r=1$.

In Figure 5, we plot the citation number versus rank for the first 10838
chemists [39], and compare this plot with present model considering $%
N_{0}=180$ and $\theta =2.5.10^{-5}$ and $i=1$ In case of Tsallis statistics
we use $N_{0}=190$ and $\mu _{r}=3.2.10^{-4}$ and $r=1$. The other
parameters are the same as in Figure 4.

The values of $N_{0}$ show that citations of Brazilian chemists is roughly $%
0.4\%$ of the total citations. The value of $\theta $ changes from $%
2.2.10^{-4}$ to $2.5.10^{-5}$. The adjustment is good both for ours and
Tsallis model.

We found that our approach considering the softening of long range
interactions, as well as Tsallis approach considering cross-over behavior
(Equation 9), gives almost the same results and can explain the entire
empirical curve including deviations for small and very large steps. Thus,
the model presented in this work is an alternative approach. The present
approach is interesting as parameter $\theta $ is related to size
limitations of the system and thus can be more informative.

\bigskip

\bigskip

\bigskip \textbf{IV. Distribution of an Entrance Examination Scores}

\bigskip

Recently we studied the statistical distribution of the student's
performance, which is measured through their marks, in the university
entrance examination (Vestibular) of UNESP (Universidade Estadual Paulista)
in Brazil, for the years 1998, 1999 and 2000. To our surprise, we observed
long ubiquitous power law tails in place of normal distribution in physical
and biological sciences [29, 35]. In humanities we have almost normal
distribution. These power law tails in physical and biological sciences
exist independently of economical, teaching, and study conditions [29]. This
shows that the power law tails are due to the nature of the subject itself.
These observations are interesting as they treats education as a complex
system and bring out the relative importance of the different factors on
science and mathematics education at high school level, which is today, an
issue of great concern in our society.

In our earlier works [29], we took marks of the students in a block of
subjects, i.e. physics, chemistry and mathematics together for physical
sciences and physics, chemistry and biology together for biological
sciences. Thus it is not possible to make a detailed quantitative analysis.
Further these are optional subjects for the examination. Thus the student's
interest for a particular area is also an important factor.

To confirm our observations, in the present work, we analyze the statistical
distribution of the marks obtained by students in individual subjects i.e.
in Physics, Mathematics, and Portuguese as native language in the Air Force
Academy entrance examination in Brazil. These students don't have any
special interest for any of these subjects as they like to have a military
career in the Air Force. Thus the statistical distribution can give better
information about the peculiar nature of each subject at high school level.

Physics and Mathematics are areas of systematic study and depend much on
regular study. To understand a chapter, the students need to know the
material given in the earlier chapters. A student who understands well the
first chapter has better conditions to understand the second and subsequent
chapters. The one, who didn't understand the first chapter, will find many
difficulties in understanding the subsequent chapters. This gives a kind of
positive feedback or long term memory effect, the reason behind power law
[8, 31]. In case of Portuguese, being a native language, each chapter is
more or less independent. Further being native language they learn
Portuguese in a natural way. Although in Portuguese there is also some
dependence of understanding earlier materials, it is not as strong as in
Physics and Mathematics.

\[
\FRAME{itbpF}{14.0518cm}{11.0424cm}{0cm}{}{}{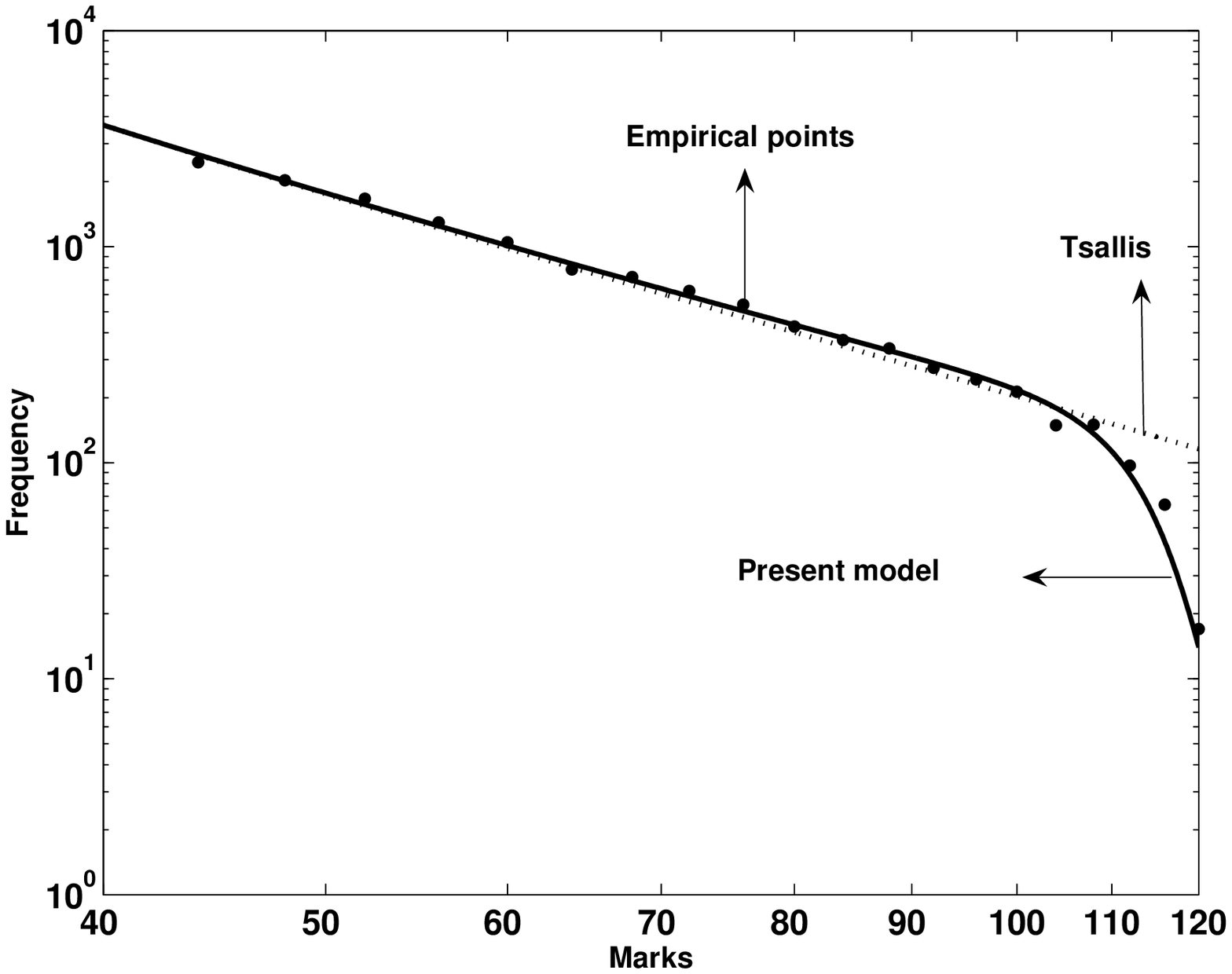}{\special%
{language "Scientific Word";type "GRAPHIC";maintain-aspect-ratio
TRUE;display "FULL";valid_file "F";width 14.0518cm;height 11.0424cm;depth
0cm;original-width 7.0828in;original-height 5.591in;cropleft "0";croptop
"0.9998";cropright "1.0058";cropbottom "0";filename
'fig6_fis0306.eps';file-properties "XNPEU";}}
\]

\begin{center}
Figure 6 -- Distribution of marks obtained by students in log-log scale in
Physics for years 2003 to 2006 considering together.
\end{center}

We compare the marks obtained by the students in Physics, Mathematics, and
Portuguese as native language. To show clearly the validity of power law we
plotted log (frequency) vs. log (marks). In figure 6, we compare the
distribution for Physics with present and Tsallis approach for all the years
2003 to 2006 together to have a better idea of the distribution. The
distribution for an individual year, i.e. 2003, 2004, 2005 or 2006 is almost
the same as all the years together. The parameters of distribution are: $%
N_{0}=4350;$ $\lambda =0.09$ and $q_{0}=1.4$. $\theta =1.265.10^{-2}$ and $%
i=12$ for our model and $%
\mu
_{r}=0.006$, and $r=1$ for Tsallis model. In figure 7, we did the same for
Mathematics. The parameters of distribution are: $N_{0}=5250$; $\lambda =0.1$%
; and $q_{0}=1.3$. $\theta =1.265.10^{-2}$ and $i=9$ for our model and $%
\mu
_{r}=0.01$, and $r=1$ for Tsallis model. We shifted the origin axis to $x_{m}
$, i.e. we use $(x-x_{m})$ in place of $x$ both in Equations (9) and (12),
where $x_{m}$ is the mark for maximum frequency. We took $x_{m}=38$ for
Physics and $39$ for Mathematics.\ We observe that the distribution in
cut-off region is better given through present approach. In figure 8, we
compare the distribution of marks obtained in Portuguese with Normal
distribution and found that Normal distribution explains the distribution
satisfactory. The parameters for Normal distribution are: $N_{0}=148700$, $%
\mu
=52.64$ and $\sigma =26.19$..

\begin{center}
\bigskip

\[
\FRAME{itbpF}{13.0479cm}{10.0364cm}{0cm}{}{}{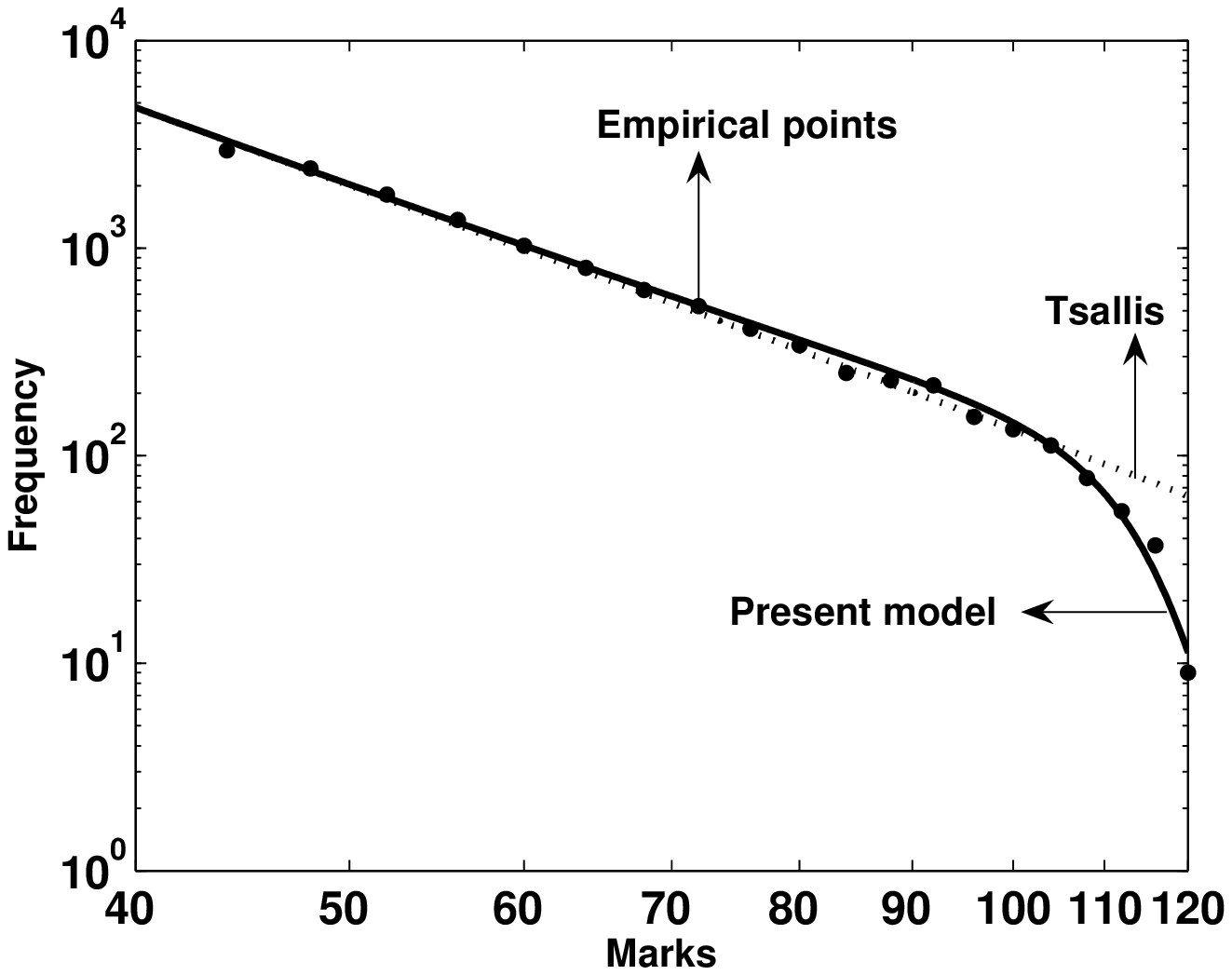}{\special%
{language "Scientific Word";type "GRAPHIC";maintain-aspect-ratio
TRUE;display "USEDEF";valid_file "F";width 13.0479cm;height 10.0364cm;depth
0cm;original-width 5.8219in;original-height 4.3708in;cropleft "0";croptop
"0.9494";cropright "0.9283";cropbottom "0";filename
'fig7_mat0306.eps';file-properties "XNPEU";}}
\]

Figure 7 -- Distribution of marks obtained by students in log-log scale in
Mathematics for years 2003 to 2006 considering together.
\end{center}

\bigskip

\bigskip

\[
\FRAME{itbpF}{13.0479cm}{10.0364cm}{0cm}{}{}{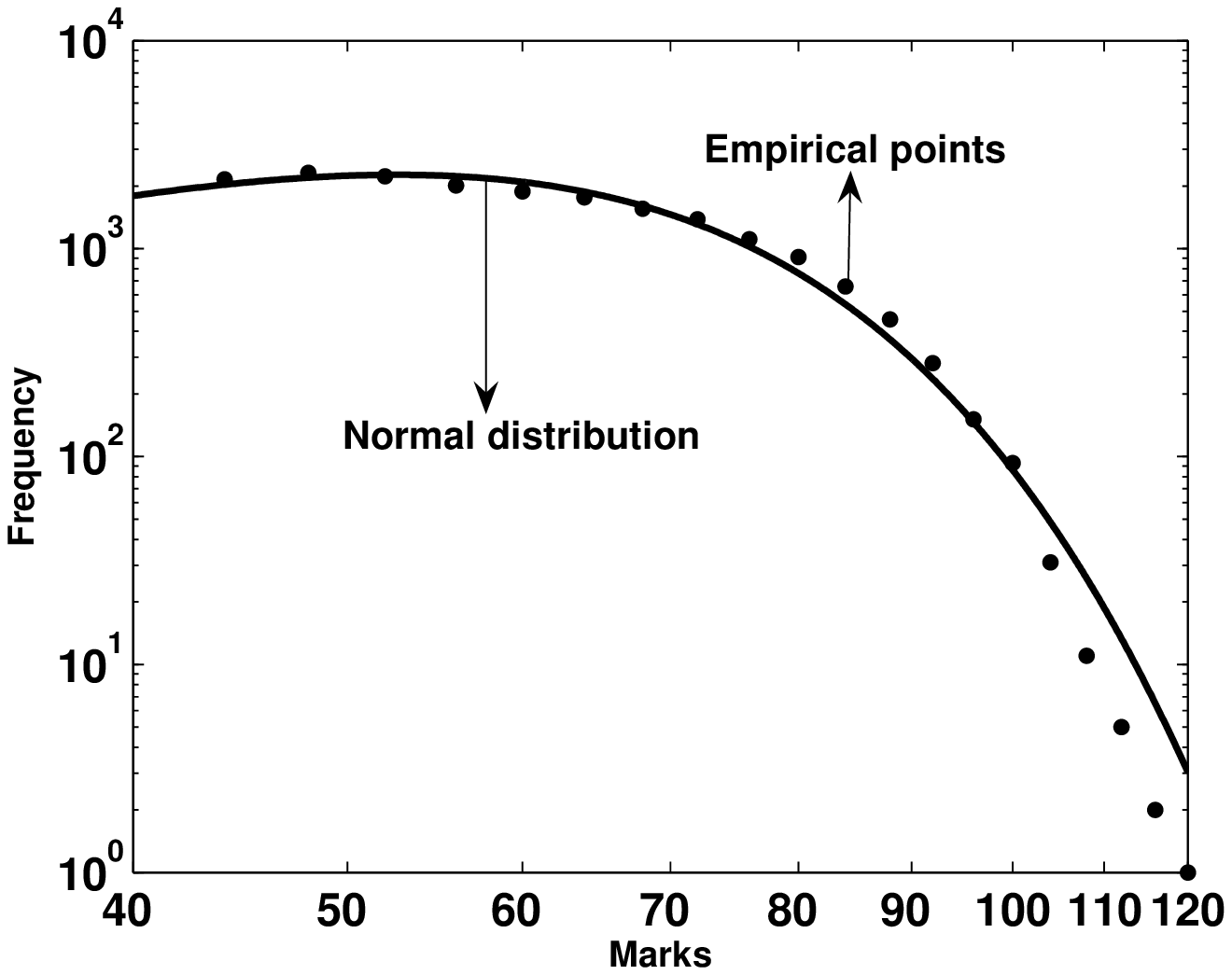}{%
\special{language "Scientific Word";type "GRAPHIC";maintain-aspect-ratio
TRUE;display "FULL";valid_file "F";width 13.0479cm;height 10.0364cm;depth
0cm;original-width 5.8219in;original-height 4.3708in;cropleft "0";croptop
"0.9494";cropright "0.9283";cropbottom "0";filename
'fig8_portugues_0306.eps';file-properties "XNPEU";}}
\]

\begin{center}
Figure 8 - Distribution of marks obtained by students in Portuguese as
native language in log-log scale for years 2003 to 2006 considering together.
\end{center}

\bigskip

\textbf{V. Discussion}

\bigskip

In case of scientist%
\'{}%
s citation index, there is no visible limit and therefore the cut-off is
slower and can be explained both through Tsallis or present approach. In
case of examination score there is a visible limit. No one can get more than
the maximum marks. This make size limiting very strong and gives a sharp
cut-off. As size limiting exists in all real systems, the present approach
is appropriate.

In conclusion, in the present paper, we presented a statistical distribution
considering that the entropic index ($q-1$), which gives information about
long range interactions and/or memory effects, decreases with step size.
This distribution gives a power law in the central part and deviates for
very small and very large steps as really observed in most of the complex
systems and thus can explain the entire distribution. This distribution is
interesting as it eliminates the necessity of truncating power law
phenomenologically [33].

\medskip

Acknowledgments

\bigskip

We are thankful to an anonymous referee for useful suggestions

.

\end{document}